\documentclass[preprint2]{aastex}

\def\hub{\ifmmode H_\circ\else H$_\circ$\fi}

\begin{document}

\title{Radial Color Gradients in K+A Galaxies in Distant Clusters of Galaxies}

\author{Lindsay J. Bartholomew, James A. Rose, Alejandro E. Gaba}
\affil{Dept. of Physics and Astronomy, University of North Carolina, Chapel Hill, NC 27599}
\email{bartholo@physics.unc.edu, jim@physics.unc.edu, gaba@physics.unc.edu} 

\author{Nelson Caldwell}
\affil{F. L. Whipple Observatory, Smithsonian Institution, P.O.Box 97, Amado, AZ 85645}
\email{caldwell@flwo99.sao.arizona.edu}

\begin{abstract}

Galaxies in rich clusters with z~$\gtrsim$~0.3 are observed to have a higher 
fraction of photometrically blue galaxies than their nearby counterparts. 
This raises the important question of what environmental effects can cause 
the termination of star formation between z~$\thickapprox$~0.3 and the 
present. The star formation may be truncated due to ram-pressure stripping, 
or the gas in the disk may be depleted by an episode of star formation caused 
by some external perturbation. To help resolve this issue, surface photometry 
was carried out for a total of 70 early-type galaxies in the cluster 
Cl1358+62, at z~$\thicksim$~0.33, using two-color images from the Hubble 
Archive. The galaxies were divided into two categories based on spectroscopic 
criteria: 24 are type K+A (e.g., strong Balmer lines, with no visible 
emission lines), while the remaining 46 are in the control sample with normal 
spectra. Radial color profiles were produced to see if the K+A galaxies show 
bluer nuclei in relation to their surrounding disks. Specifically, a linear 
gradient was fit to the radial color profile of each galaxy. We find that the 
K+A galaxies on average tend to have slightly bluer gradients towards the 
center than the normals. A Kolmogorov-Smirnov two-sample test has been 
applied to the two sets of color gradients. The result of the test indicates 
that there is only a $\thicksim$2\% probability that the K+A and normal 
samples are drawn from the same parent distribution. There is a possible 
complication from a trend in the apparent magnitude vs. color gradient 
relation, but overall our results favor the centralized star formation 
scenario as an important process in the evolution of galaxies in dense 
clusters.

\end{abstract}


\section{Introduction}

Butcher \& Oemler (1978, 1984) were the first to establish a rapid decline in
the star formation rate in rich clusters of galaxies since 
z~$\thickapprox$~0.5, when they discovered a significant increase of the blue 
galaxy fraction with increasing redshift, from about 3\% in nearby clusters, 
to 25\% in clusters at z~$\thickapprox$~0.5.  In addition, {\it Hubble Space 
Telescope} ({\it HST}) images have shown that the fraction of spirals, as 
well as interacting or merging galaxies, increases with redshift, and that 
the increase in the spiral population towards higher redshift corresponds to 
a decrease in the S0 population (Dressler et al. 1997; Fasano et al. 2000).  
Given that both gas removal, leading to globally truncated star formation 
(Gunn \& Gott 1972), and starburst-provoking (through various tidal 
perturbation scenarios, e.g., Moore et al. 1996; Bekki 1999) mechanisms are 
almost certainly operating in the rich cluster environment, it is a 
non-trivial matter to disentangle which mechanism primarily dominates galaxy 
evolution in various cluster environments.  The situation is further 
exacerbated by the fact that, except for immediately (e.g., $<$0.5 Gyr)
after a burst of star formation, which produces extremely blue colors and
strong Balmer absorption lines, a fading starburst and a fading, truncated 
disk are basically impossible to distinguish from global colors and 
spectroscopy (e.g., Newberry, Boroson, \& Kirshner 1990). Consequently, there 
has been considerable disagreement over the interpretation of the strong 
Balmer line galaxies in z~$\sim$~0.5 clusters, originally discovered by 
Dressler \& Gunn (1983), as to whether they represent post-starburst galaxies 
(Dressler \& Gunn 1983; Couch \& Sharples 1987; Barger et al. 1996; 
Dressler et al. 1999; Poggianti et al. 1999), or galaxies with globally 
truncated star formation (e.g., Balogh et al. 1999). These issues are more 
extensively summarized in Rose et al. (2001), and here we simply restate two 
key, related questions which need to be answered: (1) What form does the 
rapid cluster galaxy evolution take: is it predominantly driven by gas 
removal (Balogh et al. 1999) or a final burst of star formation 
(e.g., Dressler et al. 1999; Poggianti et al. 1999), and (2) What is the 
mechanism for the evolution: ram-pressure stripping (Gunn \& Gott 1972; 
Abadi, Moore, \& Bower 1999) or tidal perturbations (e.g., Moore et al. 1996; 
Bekki 1999)?  

As is discussed in Rose et al. (2001), ram pressure stripping of gas in 
galaxies is expected to lead to a roughly uniform truncation of star 
formation across the disk, while tidal mechanisms produce centralized bursts 
of star formation (Mihos \& Hernquist 1996). Consequently, observing the 
spatial distribution of the last star formation in galaxies can provide key 
evidence regarding the method through which the star formation has been 
extinguished.  In Rose et al. (2001) recent star formation was investigated 
in a number of early-type galaxies in nearby clusters, and it was concluded 
that this star formation is centralized. The evidence for centralized star 
formation is based on the spatial distribution of emission lines in the cases 
where star formation is still ongoing, and on the radial color profiles in 
the cases where star formation has been recently terminated. Consequently, it 
is of interest to assess the situation in clusters at higher redshift, where 
the ongoing star formation activity is at a substantially higher level. 
Recently, van Dokkum et al. (1998; hereafter VD98) completed an extensive 
study of the z~$\thicksim$~0.33 cluster CL1358+62, and have published radial 
color profiles for several galaxies, including a few in which a blue nuclear 
region is evident.  Based on these results, we have used the archival 
{\it HST} imaging data for CL1358+62, in conjunction with spectroscopic 
information published in Fisher et al. (1998; hereafter F98), to assess 
whether the cluster galaxies which show evidence of recently terminated star 
formation also have centrally concentrated blue colors.

In section~2, the spectroscopic and imaging data are summarized, as well as 
the definitions of two galaxy samples, a ``normal'' sample and a ``K+A''
sample, the latter consisting of galaxies with spectroscopic evidence of 
recently concluded star formation. In addition, the analysis techniques for 
obtaining radial color profiles are discussed. In section~3, a statistical 
comparison of the radial color profiles for the two galaxy samples is given, 
to test whether the K+A sample on average shows bluer central colors than the 
control sample. In section~4, we discuss our results within the context of 
the question of whether tidal perturbations or ram pressure stripping have 
had a greater effect on the evolution of galaxies in clusters. Throughout the 
paper we assume that \hub=70 km~s$^{-1}$~Mpc$^{-1}$ and q$_o$~=~0.15.

\section{Data Samples}

The imaging used for this paper was collected from the {\it Hubble Space
Telescope} Archive, and was combined with spectroscopic data obtained from 
the literature, for the purpose of observing the spatial distribution of 
recent star formation within cluster galaxies. The {\it HST} Archive has been
searched for galaxy clusters of moderate (e.g., $z~\sim~0.3~-~0.5$) redshift 
that have been imaged in at least two passbands, using the WFPC2 instrument. 
Our search resulted in a sample of four clusters: Abell 2390 at 
z~$\thicksim$~0.23, CL1358+62 at z~$\thicksim$~0.33, CL0303+17 at 
z~$\thicksim$~0.42, and CL0016+16 at z~$\thicksim$~0.55.  The literature on 
each cluster was then researched for spectroscopic information. The specific 
information we require is Balmer line strengths and [OII]$\lambda$3727 
emission line strengths, which are indicative of the levels of recent (e.g.,
within the past few Gyr) and current (e.g., ongoing) star formation.  

In this paper, only CL1358+62 is discussed, as the other three clusters 
were discarded for various reasons. First, they all had very limited sky 
coverage when compared to CL1358+62. The images of Abell 2390 did not include 
any of the galaxies that had been classified spectroscopically as having 
recently concluded star formation; in addition, only one exposure was taken 
at each pointing, making removal of cosmic rays difficult.  Both of the 
remaining clusters were not included in the final analysis, due to the fact 
that they had small samples of galaxies with low signal-to-noise data, which 
did not make them suitable for statistical analyses.  CL1358+62, on the other 
hand, was imaged in a mosaic over a large area, and had more than one 
exposure at each pointing, greatly facilitating cosmic ray removal.

CL1358+62 is a rich, moderate-redshift cluster, which is also X-ray luminous 
(L$_{x}$~$\thicksim$~7x10$^{44}$~ergs/s at 0.2~--~4.5 keV) (Fabricant, 
McClintock, \& Bautz 1991).  The presence of a pervasive hot intracluster 
medium along with the high velocity dispersion of the galaxies 
($\sigma$=1037~km~s$^{-1}$; F98) ensures that ram pressure stripping of the 
interstellar medium in galaxies should play an important role in the 
evolution of galaxies passing near the cluster center. F98 have carried out 
spectroscopy in CL1358+62, including line strengths for the Balmer absorption 
lines H$\beta$, H$\gamma$, and H$\delta$, and also the emission line [OII] at 
3727\AA.  

We have divided many of the member galaxies in CL1358+62 into two samples, 
based on their spectroscopic features as given in F98. One is a control 
sample, consisting of galaxies with normal spectra for old stellar 
populations, e.g., no emission and weak Balmer absorption lines. The second 
sample consists of galaxies with low emission and strong Balmer absorption. 
Originally known as E+A galaxies (Dressler \& Gunn 1983), they are now 
commonly referred to as ``K+A'' or ``k+a/a+k'' (Franx 1993; Dressler et al. 
1999; Poggianti et al. 
1999), hence we adopt the K+A terminology. Specifically, for the K+A sample, 
the average of the three Balmer lines ($<$H$>$) is required to have an 
equivalent width of at least 3\AA, and the emission strength is limited to 
less than 5\AA. The normal sample is defined to have a limit of 1.5\AA \ in 
$<$H$>$, and the [OII] line was restricted to lie between --5\AA \ and 
5\AA. F98 define normal and post-starburst galaxies slightly differently. 
Our definitions make the limits more stringent for the normal red 
absorption-line galaxy population, and slightly increase the number of 
galaxies classified as K+A. Our characterization of K+A galaxies using 
$<$H$>$ (following F98) differs from the characterization of Dressler et al. 
(1999), who used solely H$\delta$ to define their k+a/a+k galaxies. Despite 
the difference in these criteria, we have found that the H$\delta$ cut-off of 
our K+A sample is virtually identical to the 3\AA \ equivalent width 
H$\delta$ cut-off for k+a galaxies used by Dressler et al. (1999) (only three 
of our 24 K+A galaxies have H$\delta$~$<$~3\AA).

\begin{figure}
\plotone{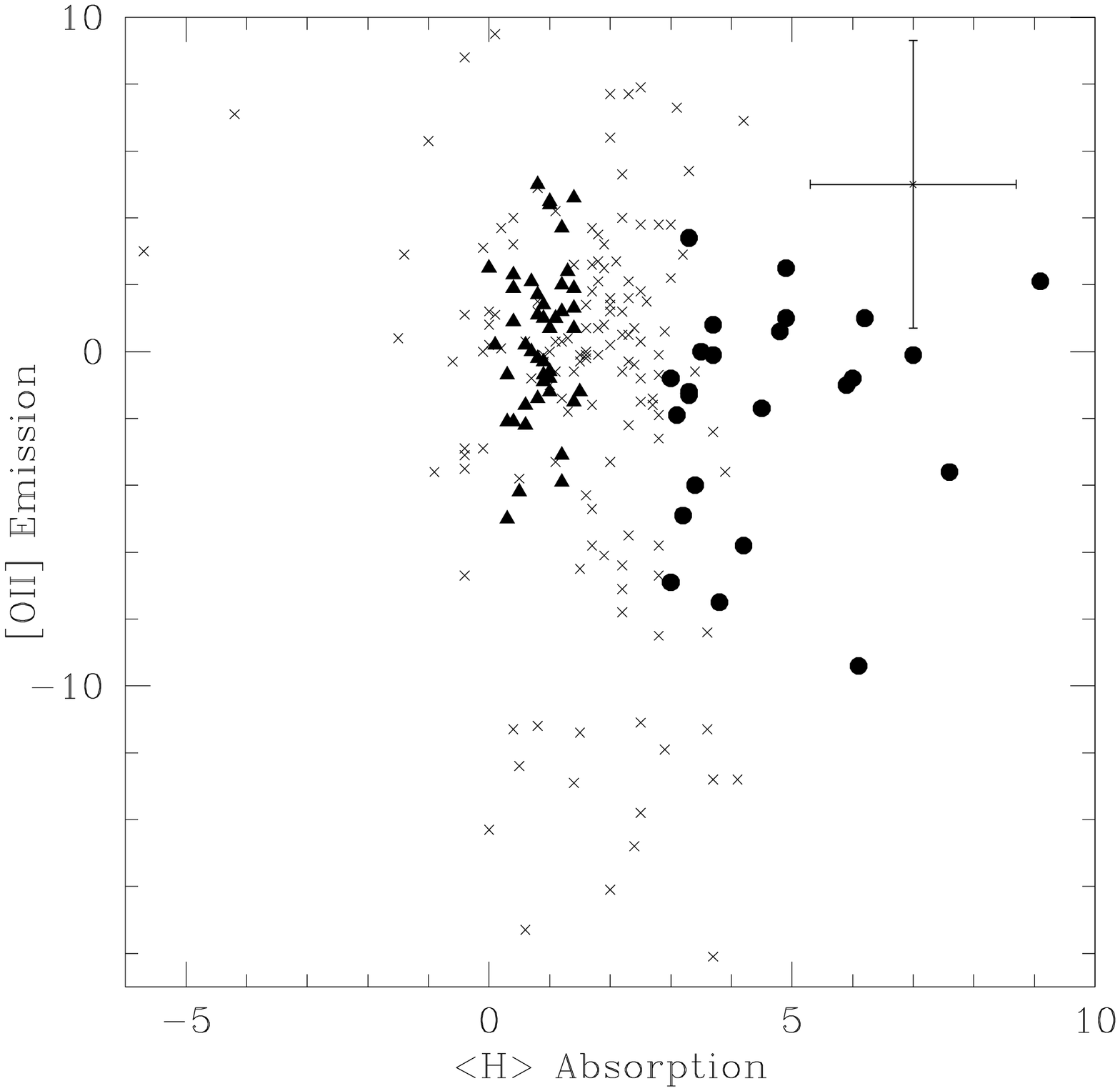}
\caption{Normal and K+A sample definitions. Filled circles are type K+A, 
filled triangles are normal type, and crosses are the rest of the sample from 
F98. Typical error bars from F98 are shown.
\label {OII}}
\end{figure}

Our final samples include 24 K+A galaxies, and 46 normal galaxies.  We have 
defined the samples so that there is a gap between them, to account for 
leakage through the large error bars. Given the rather large error bars in 
Balmer line and [OII] equivalent widths, there may be some leakage of 
galaxies from one sample into another, which would tend to weaken any 
intrinsic differences that we observe between the two samples. Fig.~\ref{OII} 
illustrates where the two samples fall on an [OII] emission vs. $<$H$>$ 
absorption plot, and shows typical error bars for the galaxies in the two 
samples as given in F98. There are some galaxies that fit within our limits 
but were excluded from the two final samples either because they were very 
close to the edge of a frame or a nearby neighbor or they could not be 
located on the images. 

We did not use any morphological criteria when originally defining our 
samples. However, VD98 have listed morphological types for the galaxies in 
the cluster, and the vast majority included in our samples (nearly 75\% in 
the K+A group, and approximately 90\% in the normal group) are type E, E/S0, 
or, more commonly, S0. The rest they have assigned ``?'' morphologies, and 
one is classified as an irregular. Three unique cases having peculiar 
morphologies will be discussed in more detail in section~3.

\subsection{Data Reductions}

Surface photometry of the galaxies in the normal and K+A samples has been 
carried out using the WFPC2 images from the {\it HST} Archive, which are 
fully described in VD98. VD98 imaged the cluster at 12 different positions in 
two filters (F606W at 5957\AA \ and F814W at 7940\AA), creating a mosaic 
with a total area covering 49 arcmin$^{2}$. Three 1200s exposures were taken 
at each position, for each filter. Image analysis has been carried out using 
the IRAF package. The first step is to combine the images at each position to 
remove cosmic rays, using the `imcombine' routine. Cosmic rays near a galaxy 
can also be removed via the `imedit' task. A general sky-subtraction is 
accomplished by statistically computing the average sky value in an area void 
of galaxies, and subtracting that value from the entire image. VD98 have 
described large-scale diagonal bands in the background of some of the images, 
produced by light scattered off the Optical Telescope Assembly. We avoided 
attempting surface photometry for any galaxies which are located in an area 
whose background varies too greatly because of this effect.

We have located each of the galaxies in the two samples, using the x and y 
offset coordinates from the cluster center as given in Table~1 of 
F98.\footnote{We were able to identify the galaxies in the VD98 images only 
if we used for the cluster center RA=13$^{h}$58$^{m}$21$\fs$0 and 
Dec=62$\arcdeg$45$\arcmin$37$\arcsec$, B1950, instead of the central 
coordinates given in F98.} To construct the radial color profiles for each 
galaxy, the `ellipse' routine in IRAF has been used. The `ellipse' task fits 
azimuthally--averaged elliptical isophotes to a galaxy as a function of its 
semi--major axis. The `ellipse' routine is run for each galaxy in both the 
F606W and F814W images. Combining the F606W and F814W magnitude profiles 
results in a radial color profile of the galaxy. A linear least squares fit 
is made to the F606W--F814W color (which can be transformed to V--I color, 
and which, at z~$\thicksim$~0.33, corresponds to a rest frame color between 
B--V and B--R) versus the logarithm of the semi--major 
axis.\footnote{Hereafter the logarithmic color gradient d(F606W--F814W)/dlogR 
is referred to as the color index CI.} The zero-point of the scale is 
arbitrary, and error bars are included in the profiles, which includes both 
statistical errors (calculated for the magnitudes by `ellipse' from the 
scatter in the intensity profile of the fit) and sky subtraction errors (we 
estimate a 1\% error in the sky background), which become important at large 
radii. 

The differing point spread function for the F606W and F814W images proves to 
be a problem for the inner $\thicksim$1.5 pixels. While we degraded the F814W 
frame to match the F606W frame, the color profile is not reliable within that 
radius, hence we only consider each profile outside of $\thicksim$1.5 pixels. 
For CL1358+62, excluding 1.5 pixels corresponds to losing the inner 0.66~kpc 
of each galaxy. Since K+A galaxies in nearby clusters are observed to have 
recent star formation typically extending through the central 2~kpc in radius
(Caldwell et al. 1996; Caldwell, Rose, \& Dendy 1999; Rose et al. 2001), 
we might expect a similar radius for such activity in the z~$\thicksim$~0.3 
galaxies. Thus the inner cut-off, while producing a significant loss in the 
lever arm for measuring color gradients in the central regions, still allows 
for the detection of centralized star formation over the spatial scales 
expected from nearby examples. In addition, to keep a consistent comparison 
standard for the entire sample, the fit for each galaxy was restricted to a 
radius of 0.7$\arcsec$ ($\thicksim$7 pixels), which corresponds to roughly 
3.1~kpc for CL1358+62. This limit was chosen because it is large enough to 
encompass a centralized burst of the kind seen in nearby K+A galaxies, but 
small enough so as not to include a significant amount of area in which the 
sky subtraction becomes overly important.

\section{Results}

\begin{figure}
\plotone{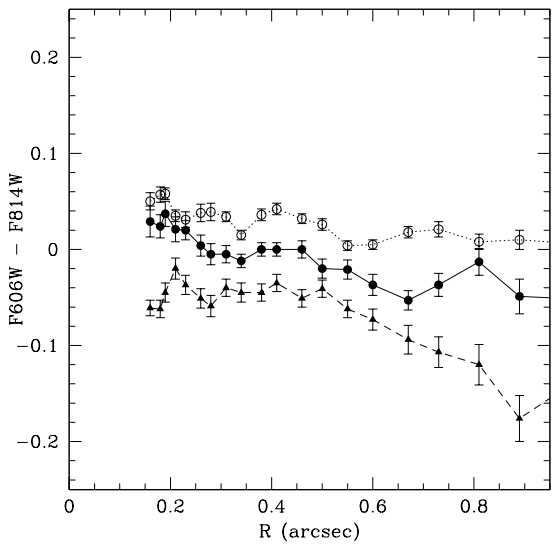}
\caption{Examples of color profiles in the normal sample. They are mainly 
flat, with a slightly negative slope. The filled circles represent galaxy 
\#236, the open circles galaxy \#269, and the triangles galaxy \#347. The 
zero-point is arbitrary. 
\label{normprof}}
\end{figure}

\begin{figure}
\plotone{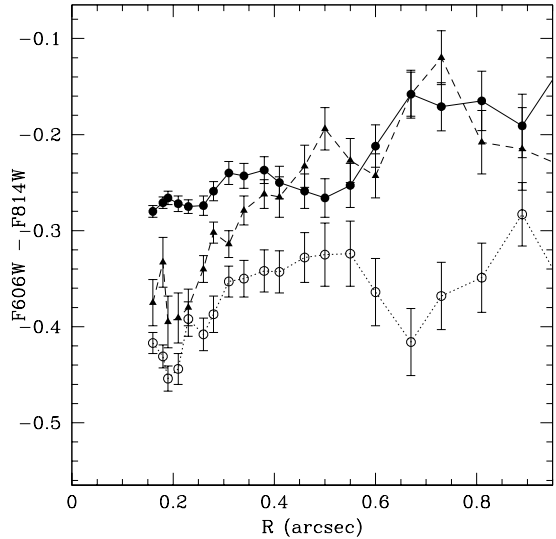}
\caption{Examples of color profiles in the K+A sample. The profiles with 
triangular points (\#562) and open circles (\#420) are in good agreement with 
the profiles given for these galaxies in VD98. The filled circles represent 
galaxy \#226. The zero-point is arbitrary.
\label{eaprof}}
\end{figure}

Examples of color gradients for both the normal and K+A samples are shown in 
Figs. \ref{normprof} and \ref{eaprof}. As a check on our methods, we have 
qualitatively compared our results for 4 of the 6 galaxies (\#109, 211, 420, 
\& 562) whose radial color gradients are plotted in Fig.~12 of VD98, and find 
good agreement. The results of the least squares linear fit to the radial 
color gradient for the normal sample are given in Table~1, where the 
logarithmic color gradient in F606W--F814W is listed for each galaxy in 
column~9, along with the galaxy ID (col.~1), x--y coordinates in arcseconds 
from cluster center (cols.~2 and 3), apparent V magnitude and morphological 
classification, (cols.~4 and 5), and [OII]$\lambda$3727 line strength and 
average Balmer absorption strength $<$H$>$ (cols.~6 and 7). The data in cols. 
1--7 are taken from VD98 and F98; the rest-frame absolute B magnitudes are 
given in column~8. Table~2 shows similar information for the K+A sample.

\begin{figure}
\plotone{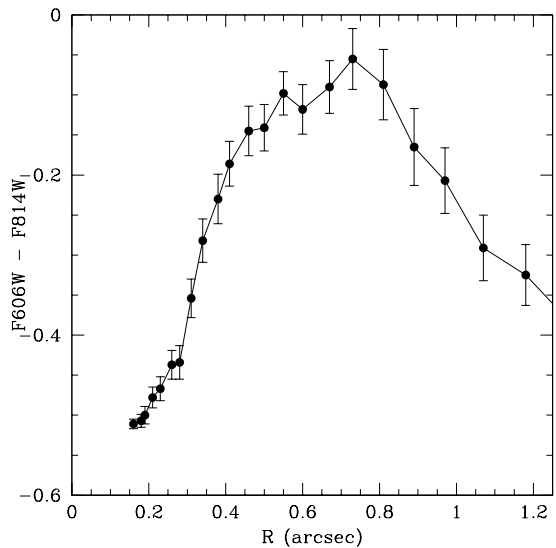}
\caption{Example of an extreme radial color profile of a K+A galaxy, \#200,
which deviates from a simple linear color gradient.
\label{g200prof}}
\end{figure}

\begin{figure}
\plotone{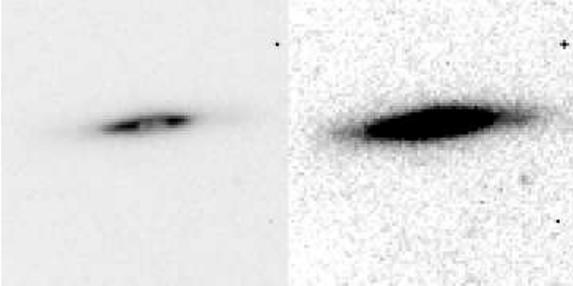}
\caption{F606W image of galaxy \#200 in Fig.~\ref{g200prof}, which has the 
odd radial color profile. Two contrast levels are shown, to reveal its 
peculiar morphology. 
\label{g200im}}
\end{figure}

\begin{figure}
\plotone{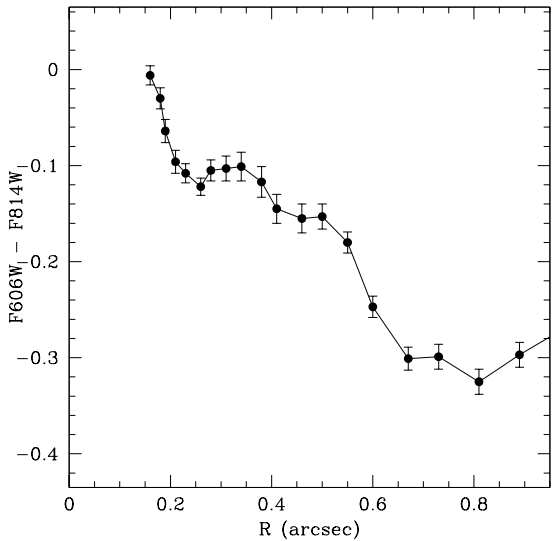}
\caption{A K+A galaxy, \#507, with an extreme red nucleus. 
\label{g507prof}}
\end{figure}

\begin{figure}
\plotone{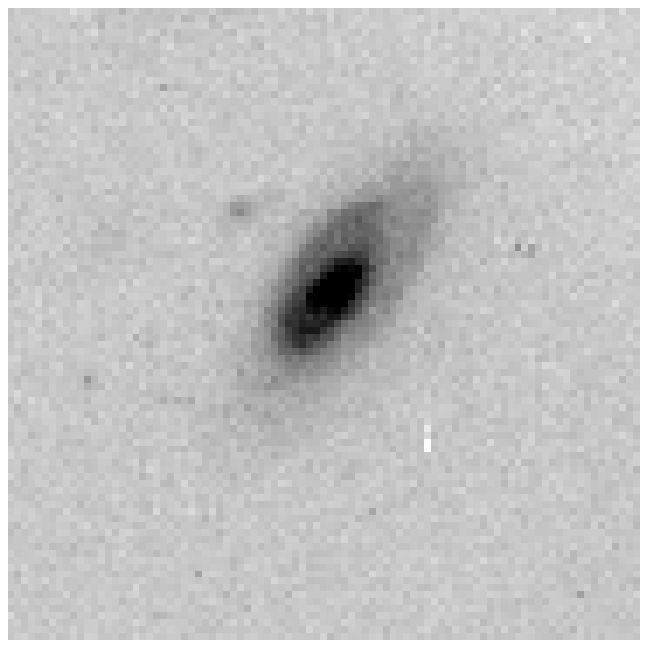}
\caption{F606W image of galaxy \#507 in Fig.~\ref{g507prof} with the extreme 
red nucleus. The galaxy has a disturbed morphology, with an asymmetric spiral 
arm. 
\label{g507im}}
\end{figure}

As stated earlier, most galaxies in the two samples have normal morphologies 
for early-type galaxies. With the exception of one galaxy classified as 
irregular (\#290), and two assigned an S0 morphology (\#200 \& \#507) (VD98), 
the galaxy morphologies are sufficiently regular that their radial profiles 
can be readily fit with the `ellipse' routine. These three special cases (all 
in the K+A sample) have morphological peculiarities which would preclude a 
reliable fit with elliptical isophotes. Fig.~\ref{g200prof} shows the radial 
color profile of galaxy \#200, which is not well fit by a simple linear color 
gradient with radius.  An image of this galaxy is shown in Fig.~\ref{g200im}, 
and reveals serious morphological peculiarities. Galaxy \#200 is evidently a 
nearly edge-on disk galaxy, which may cause a problem with the 
ellipse-fitting technique.  In addition, when viewed at high contrast (see 
Fig.~\ref{g200im}),  strong deviations from a normal disk galaxy radial 
luminosity profile are evident. A second special case in our K+A sample, 
\#507, has an extreme red nucleus and morphological irregularities.  Despite 
the red nucleus, \#507 does have relatively strong Balmer line strengths (and 
is classified in F98 as an E+A galaxy), but it is morphologically disturbed, 
in that it has a single asymmetric spiral feature, which produces problems 
for the ellipse-fitting technique. Similar morphological disturbances in 
cluster galaxies with strong Balmer lines have been found by Dressler et al. 
(1999). The radial color profile and image of \#507 are shown in 
Figs. \ref{g507prof} and \ref{g507im}. Due to the morphological peculiarities 
of these two galaxies (as well as the irregular galaxy), the final 
statistical tests have been repeated first including, then excluding them, to 
ensure they are not contaminating the results.

\begin{figure}
\plotone{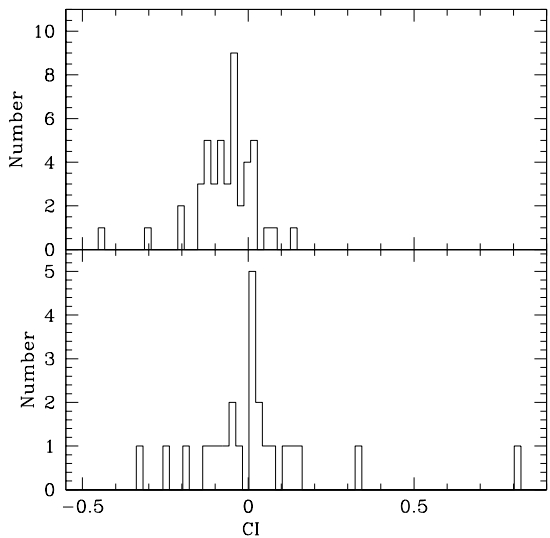}
\caption{Histogram of radial color gradients. The top panel is the 
normal sample, the bottom panel the K+A's. The color gradient (CI) is defined 
as d(F606W--F814W)/dlogR, where R is the semi-major axis in arcsec. The 
galaxy to the far right in the K+A group is \#200, and \#507 is at the far 
left. 
\label{histuncorr}}
\end{figure}

\begin{figure}
\plotone{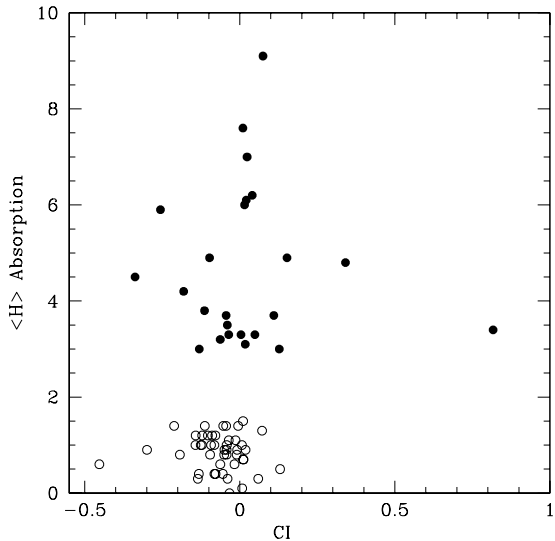}
\caption{Balmer line strength is plotted versus the observed radial color 
gradients (CI).  Filled circles are for K+A galaxies, open circles are
for normal galaxies. The filled circle to the far right is galaxy \#200; the 
filled circle to the far left is \#507.
\label{aveH}}
\end{figure}

To compare the color gradients of the K+A and normal samples, we have
constructed histograms of the two samples, shown in Fig.~\ref{histuncorr}. 
The histograms show the entire samples of galaxies, including the peculiar 
cases discussed specifically above. Our normal sample has an average slope of 
d(F606W--F814W)/dlogR~$\thicksim$~--0.07 (where R is the semi-major axis in 
arcsec). For reference, Peletier et al. (1990) found that typical early-type 
galaxies have color gradients of d(B-R)/dlogR~$\thicksim$~--0.1. This lends 
confidence to our results, since at the redshift of CL1358+62, the F606W band 
is close to the rest wavelength B band, and the F814W band is between rest 
wavelength V and R. The histogram for the K+A sample, excluding the specific 
cases discussed above, peaks at a more positive slope than the normal sample, 
at very nearly zero. Fig.~\ref{aveH} illustrates the scatter plot of Balmer 
line strength versus the color gradient. The K+A galaxies with Balmer line 
strengths between 3 and 4\AA \ (that were not included in the E+A category 
of F98) lie in the middle of the gradient distribution, yet still displaced 
from the normal group. While the overall shift between K+A and normal 
galaxies is evident, no monotonic trend of Balmer line strength with color 
gradient is apparent, although above 6\AA \ in equivalent width, all slopes 
are positive. 

\begin{figure}
\plotone{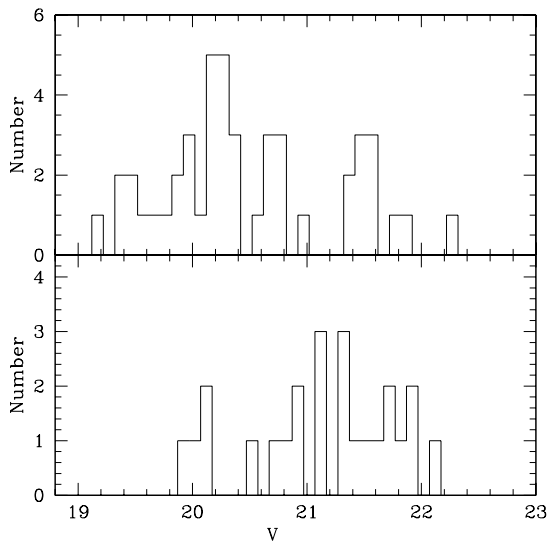}
\caption{The apparent V magnitude distributions for the normal 
(top) and K+A (bottom) samples. 
\label{histmag}}
\end{figure}

\begin{figure}
\plotone{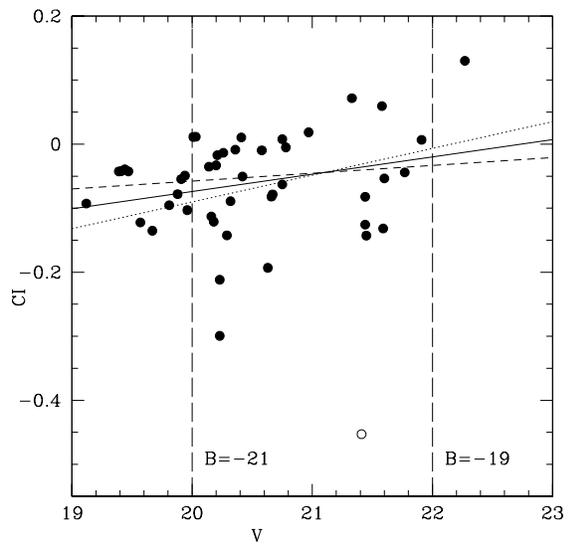}
\caption{Radial color gradients are plotted versus apparent V magnitude 
for the
normal galaxy sample.  The solid line represents a linear least squares fit,
with the open circle point excluded.  The dashed and dotted lines represent
the $\pm$1$\sigma$ uncertainties in the slope of the relation.  The
vertical dotted lines show rest--frame absolute B magnitudes.
\label{gradmag}}
\end{figure}

We applied a Kolmogorov--Smirnoff (K--S) two-sample test to find the 
probability that the two samples are drawn from the same parent distribution. 
The three galaxies with problematic morphologies were excluded from the 
statistical test.  The result of the test of the distribution of color 
gradients within the two samples (compiled based on spectroscopic criteria) 
is that there is only a 2.5\% probability that the samples are drawn from the 
same parent sample. To further test the result, we ran the K--S test, first 
including each of the previously excluded galaxies (\#290, 200, \& 507), and 
then including all three. The confidence levels ranged between 1.8--0.6\%. 

A potential complication can arise if the two samples have different
distributions in apparent magnitude, and if there is a correlation between
apparent magnitude and the radial color gradient. Using the apparent 
magnitudes listed in VD98 (transformed from the $\it{HST}$ filters to the V 
band), we see that there is a difference in the mean apparent V magnitudes of 
the two samples, a fact that is illustrated in Fig.~\ref{histmag}. In 
addition, we do find evidence for a trend in the apparent V magnitude with 
the radial color gradient.  This trend is illustrated in Fig.~\ref{gradmag}, 
where the logarithmic color gradient is plotted versus apparent V magnitude 
for the normal galaxy data.  To account for this systematic effect, a linear 
least squares fit was made to the data, and the gradients of both the normal 
galaxies and K+A galaxies were corrected to the same reference magnitude of 
V=21.1. These corrected color gradients, using the best fit slope of 
Fig.~\ref{gradmag}, are listed in column~10 of Tables~1 and 2. The linear 
fit, however, is quite uncertain, and the $\pm$1$\sigma$ slopes are also 
plotted on Fig.~\ref{gradmag}. 

After correcting the observed radial color gradients in each galaxy to a 
single reference apparent magnitude, we ran the K--S test again on the K+A 
and normal galaxy samples, to evaluate whether the two samples still appear 
to be statistically different in their color gradient distributions. We find 
that the result of the K--S test, shown in Table~3, depends strongly on 
whether the maximum, best fit, or minimum slope from Fig.~\ref{gradmag} is 
used. In col.~2 we used our raw color gradient data, and in cols.~3--5 we 
show the results after the correction for apparent magnitude, using the 
minimum, best fit, and maximum slopes shown in Fig.~\ref{gradmag}, 
respectively. Clearly, the magnitude of the correction is crucial to deciding 
whether the two samples are fundamentally different in their color gradient 
properties.

\begin{figure}
\plotone{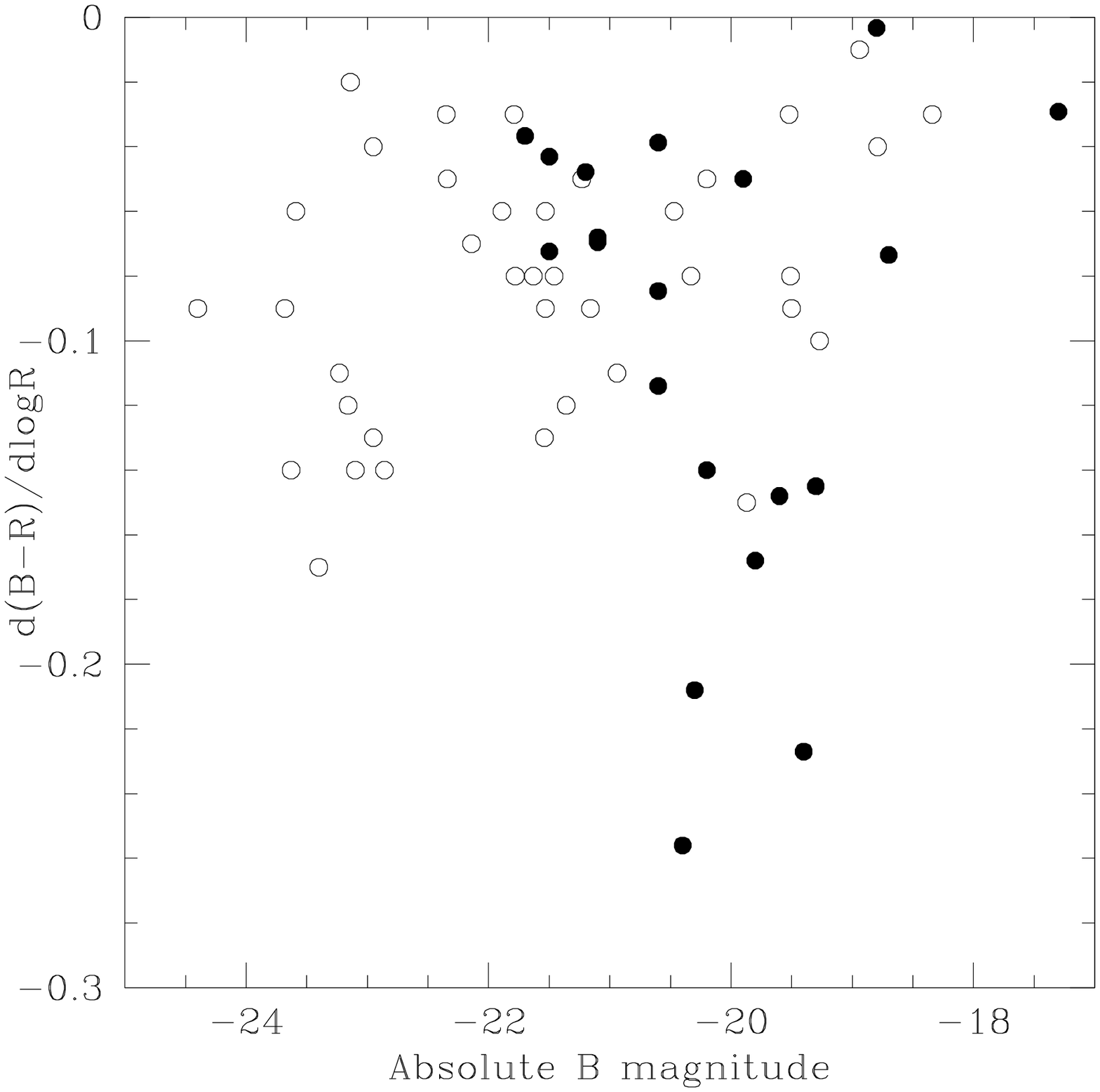}
\caption{The distribution of B-R logarithmic color gradients vs. absolute B 
magnitude for nearby early-type galaxies from Peletier et al. (1990) (open 
circles), and Vader et al. (1988) (filled circles).
\label{vadpel}}
\end{figure}

Since the observed trend between the color gradients and apparent V magnitude
in CL1358+62 is quite uncertain, and since the amount of correction for 
apparent magnitude has an important effect on the interpretation of the 
results, we have investigated other studies of radial color gradients in 
galaxies. In particular, we have assessed the literature on nearby early-type
galaxies, where the higher spatial resolution offers more reliable results.
Absolute magnitude versus color gradient plots were constructed from the 
studies of Peletier et al. (1990), Vader et al. (1988), Jansen et al. (2000), 
and Balcells \& Peletier (1994).  Fig.~\ref{vadpel} illustrates the data from 
Peletier et al. (1990) and Vader et al. (1988), where we have excluded the 
faint galaxies with $M_B>-18$.\footnote{The faintest galaxies in the Vader et 
al. (1988) sample, all Virgo cluster galaxies with $M_B>-18$, were found to 
have reverse color gradients. Recently, Concannon, Rose, \& Caldwell (2000) 
have shown that these Virgo galaxies show a large variety of Balmer 
absorption line strengths, indicative of a large variation in age.  All of 
the galaxies studied in CL1358+62 are brighter than $M_B=-18$, which can be 
seen in Fig.~\ref{gradmag}, where the vertical dashed lines show the 
locations of rest wavelength B~=~--21 and B~=~--19.  Hence we excluded the 
faint galaxies from  Vader et al. (1988) from Fig.~\ref{vadpel}.} There is no 
observable trend in color gradient with absolute magnitude.

In summary, we have observed galaxy samples in the cluster CL1358+62, at 
z~$\thicksim$~0.33. For those galaxies with strong Balmer absorption lines, 
and little emission, the radial color profiles have less of a red slope 
towards the center than typical E/S0 galaxies, in that bluer colors are more 
likely to appear in the central regions.  There is some uncertainty as to 
whether this effect can be due to a systematic trend of color gradient with 
apparent magnitude.  However, given the uncertainty in the observed trend in 
CL1358+62, along with the fact that no such trend is observable among nearby 
galaxies, we conclude that the observed difference in the color gradients 
between normal and K+A galaxies in CL1358+62 is likely due to a real 
intrinsic difference in the two samples.

\section{Discussion}

There have been a number of studies investigating the issue of how star 
formation typically ends in cluster galaxies, e.g., whether through
sudden truncation of normal star formation, or through a starburst. 
We briefly summarize these previous studies here, then place our new
result within that context.

On the one hand, evidence exists that for many galaxies in rich clusters,
rapid truncation of normal star formation is the principal evolutionary
mechanism. For example, Balogh et al. (1997, 1998) concluded from a study of 
the equivalent widths of the [OII]$\lambda$3727 emission line in cluster and 
field galaxies, that star formation is suppressed in the cluster relative 
to the field, indicating that galaxies lose their gas through truncation of 
star formation, not through a starburst.  In addition, there is abundant 
evidence that spiral galaxies in nearby clusters are on average globally 
depleted in HI when compared to their field counterparts (Solanes et al. 
2001, and references therein) and that the sizes of the HI disks have been 
reduced (Cayatte et al. 1994).  Ram pressure stripping is clearly the most 
plausible explanation for these observations. Furthermore, several studies 
have provided direct evidence for spirals in the process of losing gas due 
to ram pressure stripping, in that displaced HI envelopes and/or bow shock 
morphologies in truncated H$\alpha$ disks are observed (Gavazzi et al. 1995; 
Kenney \& Koopmann 1999; Bravo-Alfaro et al. 2000). 

On the other hand, numerical simulations by Abadi, Moore, \& Bower (1999) 
indicate that ram pressure stripping cannot be solely responsible for the 
observed rapid evolution of the star formation rate in cluster galaxies.  In 
addition, while this mechanism does lead to stripping preferentially in the 
outer parts of a galaxy disk (which may explain the group of Virgo spirals 
found by Cayatte et al. 1994 to have centrally concentrated HI disks), even 
under the most favorable case of perpendicular infall the gas disk is {\it 
not} stripped down to within only $\sim$2~kpc of the nucleus (as indicated by 
the radial extent of observed blue nuclear regions in K+A galaxies).
Furthermore, imaging of clusters out to z~=~0.83 has revealed an enhanced
level of tidally-interacting systems (Thompson 1988; Lavery \& Henry 1988;
Lavery, Pierce, \& McClure 1992; Dressler et al. 1994; Couch et al. 1998; van 
Dokkum et al. 1999). Dressler et al. (1999) and Poggianti et al. (1999) have 
also seen evidence of tidal interactions in disk galaxies, and conclude that 
a burst of star formation is more likely in these cluster galaxies than a 
truncation of normal star formation. As well, studies of nearby clusters 
reveal examples of galaxies which have experienced centralized star formation 
episodes (Moss \& Whittle 1993, 2000; Caldwell et al. 1996; Caldwell, Rose, 
\& Dendy, 1999; Rose et al. 2001). In particular, most of the 
z~$\thicksim$~0.05 K+A galaxies studied to date show strong color gradients 
indicating blue central regions. Norton et al. (2001) have concluded, based 
on long-slit spectra of a sample of field E+A galaxies identified by 
Zabludoff et al. (1996), that the young stellar populations of these galaxies 
are more centrally concentrated than the old stellar populations. 
Furthermore, Zabludoff et al. (1996) found that many of their field E+A 
galaxies have peculiar or interacting morphologies, which they argue is due 
to galaxy-galaxy interactions.  Also, Moss \& Whittle (1993, 2000) 
find that early-type tidally-distorted spirals are often found with 
compact nuclear H$\alpha$ emission, thus providing a direct link between 
tidal perturbation and centralized star formation.  Consequently, centralized 
star formation, probably induced by tidal disturbances, appears to play an 
important role in cluster evolution as well.

In our study we have found evidence for systematically different radial color
gradients between galaxies in CL1358+62 classified (on spectroscopic grounds)
as K+A and those classified as having normal older populations.  The
difference is in the sense that bluer nuclei (relative to the surrounding
area outside the central $\sim$2~kpc) prevail in the case of the K+A 
galaxies. We found this difference in spite of the fact that there could be 
some leakage between the two samples, due to the large error bars in the 
spectroscopic measurements, which would tend to blur the difference we find. 
The difference could be due to an apparent trend in the relationship between 
magnitude and color gradient, which greatly affects the statistical strength 
of our results.  However, the absence of such an observed trend in nearby 
galaxies indicates that there is likely a real systematic difference between 
the two samples.

Our result is certainly compatible with a centralized episode of star 
formation that naturally occurs in various tidal interaction scenarios, 
ranging from equal-mass merger models (Mihos \& Hernquist 1996) to galaxy 
``harrassment'' (Moore et al. 1998) and tidal interaction between a galaxy 
and the strongly varying mean cluster tidal field (Bekki 1999).  Because 
nearly all of the K+A galaxies are classified as S0's, and thus probably 
still maintain a major disk structure, it is unlikely that equal mass mergers 
can be responsible, since they tend to destroy disks (Dressler et al. 1999; 
Poggianti et al. 1999).  Thus one of the alternate tidal 
scenarios must predominate.  The bluer nuclei of the K+A galaxies does not 
appear compatible with a sudden global quenching of star formation, since 
that mechanism would be expected to leave behind a red bulge surrounded by a 
gradually fading and reddening disk, e.g., at all times producing a 
logarithmic radial color gradient that is negative, such as is seen in 
early-type S0's, e.g., NGC3115 (Silva et al. 1989).  The only way to 
reconcile the sudden quenching scenario with our observed radial color 
profiles is to have the disk eaten away from the outside in, {\it over a 
considerable period of time}.  In summary, while the high X-ray luminosity of 
CL1358+62 indicates that ram pressure stripping must play an important role 
in the evolution of its member galaxies, tidal interaction may be playing a 
key role as well.

\acknowledgements

This research has been partially supported by STScI Archival Grant 
\#HST-AR-08734.01-A to the University of North Carolina and to the 
Smithsonian Astrophysical Observatory.

\clearpage
\begin{deluxetable}{crrrrrrrrrr}
\tablecaption{Normal Galaxies \label{tbl-1}}
\tablewidth{0pt}
\tablehead{
\colhead{Gal\#} & \colhead{x} & \colhead{y} & \colhead{V} & 
\colhead{Type} & \colhead{[OII]$\lambda$3727} & \colhead{$<$H$>$} & 
\colhead{$M_B$\tablenotemark{*}} & \colhead{CI\tablenotemark{\dag}} & 
\colhead{[CI]$_c$\tablenotemark{\ddag}} 
\\
\colhead{} & \colhead{($\arcsec$)} & \colhead{($\arcsec$)} & \colhead{} & 
\colhead{} & \colhead{(\AA)} & \colhead{(\AA)} & \colhead{} & \colhead{} 
}
\startdata
108 &$-$131.10 &$-$224.99 &20.58 &E &1.7$\pm2.6$ &0.8$\pm1.3$ &$-$20.45 &$-$0.010 &0.004\\
126 &67.93 &$-$208.17 &20.01 &S0 &2.1$\pm1.8$ &0.7$\pm1.1$ &$-$21.02 &0.011 &0.040\\
129 &127.68 &$-$201.66 &19.96 &S0 &$-$3.9$\pm1.5$ &1.2$\pm0.8$ &$-$21.07 &$-$0.103 &$-$0.072\\
142 &$-$50.94 &$-$191.34 &20.26 &S0 &1.0$\pm2.8$ &1.1$\pm1.9$ &$-$20.77 &$-$0.014 &0.009\\
164 &$-$105.06 &$-$172.23 &19.88 &S0 &0.9$\pm3.0$ &0.4$\pm1.6$ &$-$21.15 &$-$0.078 &$-$0.045\\
182 &$-$177.65 &$-$152.25 &20.97 &S0 &$-$0.7$\pm3.5$ &0.9$\pm2.1$ &$-$20.06 &0.018 &0.022\\
215 &28.54 &$-$127.40 &20.78 &S0 &0.7$\pm3.7$ &1.4$\pm2.1$ &$-$20.25 &$-$0.005 &0.004\\
233 &$-$164.92 &$-$114.30 &19.47 &E &1.4$\pm1.9$ &0.9$\pm0.8$ &$-$21.56 &$-$0.043 &0.001\\
235 &$-$46.46 &$-$111.33 &21.60 &? &1.3$\pm6.5$ &1.4$\pm1.9$ &$-$19.43 &$-$0.053 &$-$0.067\\  
236 &$-$37.20 &$-$110.38 &20.16 &S0 &4.6$\pm3.2$ &1.4$\pm2.1$ &$-$20.87 &$-$0.113 &$-$0.088\\
239 &$-$26.80 &$-$103.88 &20.67 &E &3.7$\pm4.1$ &1.2$\pm1.6$ &$-$20.36 &$-$0.078 &$-$0.066\\
242 &$-$8.55 &$-$101.16 &19.81 &E &5.0$\pm2.6$ &0.8$\pm1.8$ &$-$21.22 &$-$0.095 &$-$0.060\\
248 &$-$59.75 &$-$96.18 &20.20 &E &2.5$\pm4.7$ &0.0$\pm1.2$ &$-$20.83 &$-$0.033 &$-$0.009\\
254 &$-$63.46 &$-$91.22 &19.91 &E &1.9$\pm3.6$ &0.4$\pm1.5$ &$-$21.12 &$-$0.055 &$-$0.023\\
269 &$-$16.73 &$-$76.88 &19.12 &E &0.7$\pm1.7$ &1.0$\pm1.2$ &$-$21.91 &$-$0.093 &$-$0.040\\
278 &$-$19.71 &$-$73.64 &20.66 &S0 &2.3$\pm3.4$ &0.4$\pm1.2$ &$-$20.37 &$-$0.082 &$-$0.070\\
288 &$-$75.07 &$-$68.20 &20.75 &S0 &0.2$\pm8.9$ &0.1$\pm1.9$ &$-$20.28 &0.008 &0.017\\ 
298 &$-$19.71 &$-$58.77 &19.44 &S0 &$-$0.7$\pm6.0$ &0.3$\pm1.1$ &$-$21.59 &$-$0.039 &0.006\\ 
299 &$-$45.12 &$-$54.61 &21.45 &E &4.4$\pm3.6$ &1.0$\pm1.6$ &$-$19.58 &$-$0.143 &$-$0.152\\   
303 &$-$88.96 &$-$53.79 &20.03 &E &0.0$\pm4.3$ &0.7$\pm1.2$ &$-$21.00 &0.011 &0.040\\ 
347 &$-$51.92 &$-$22.89 &20.14 &E &1.0$\pm5.6$ &1.1$\pm1.8$ &$-$20.89 &$-$0.035 &$-$0.009\\ 
353 &7.80 &$-$17.57 &19.41 &E &0.7$\pm1.8$ &1.0$\pm0.8$ &$-$21.62 &$-$0.042 &0.004\\
371 &87.25 &$-$2.90 &19.67 &S0 &$-$2.1$\pm2.0$ &0.3$\pm1.1$ &$-$21.36 &$-$0.135 &$-$0.096\\ 
394 &110.45 &20.22 &21.33 &S0 &2.4$\pm12.7$ &1.3$\pm2.1$ &$-$19.70 &0.072 &0.066\\   
408 &$-$5.33 &25.92 &20.21 &S0 &0.2$\pm2.5$ &0.6$\pm1.5$ &$-$20.82 &$-$0.017 &0.007\\   
416 &2.95 &33.74 &21.44 &E &4.5$\pm6.8$ &1.0$\pm2.0$ &$-$19.59 &$-$0.126 &$-$0.135\\   
421 &$-$32.31 &37.35 &20.75 &E &$-$1.6$\pm4.7$ &0.6$\pm1.9$ &$-$20.28 &$-$0.063 &$-$0.054\\ 
434 &1.34 &45.48 &21.59 &? &$-$2.1$\pm4.0$ &0.4$\pm1.3$ &$-$19.44 &$-$0.132 &$-$0.145\\
444 &154.20 &56.42 &20.23 &E &$-$1.5$\pm4.9$ &1.4$\pm1.3$ &$-$20.80 &$-$0.212 &$-$0.189\\ 
460 &51.45 &70.69 &21.44 &S0 &$-$1.2$\pm4.4$ &1.0$\pm1.7$ &$-$19.59 &$-$0.082 &$-$0.091\\   
468 &171.80 &83.16 &20.36 &S0 &$-$0.9$\pm3.1$ &0.9$\pm1.3$ &$-$20.67 &$-$0.009 &0.011\\
470 &131.21 &84.61 &19.57 &E &$-$0.8$\pm1.9$ &1.0$\pm1.1$ &$-$21.46 &$-$0.122 &$-$0.081\\   
523 &37.47 &126.64 &20.23 &S0 &1.0$\pm2.8$ &0.9$\pm2.0$ &$-$20.80 &$-$0.299 &$-$0.276\\
536 &17.01 &144.87 &19.39 &E &1.1$\pm1.6$ &0.8$\pm0.6$ &$-$21.64 &$-$0.043 &0.003\\  
537 &31.69 &145.16 &20.63 &E &$-$0.2$\pm1.8$ &0.8$\pm1.1$ &$-$20.40 &$-$0.193 &$-$0.180\\   
542 &$-$4.28 &150.03 &20.18 &E &1.2$\pm2.2$ &1.2$\pm0.9$ &$-$20.85 &$-$0.121 &$-$0.096\\
544 &$-$195.75 &151.37 &20.42 &S0 &$-$1.4$\pm2.3$ &0.8$\pm1.1$ &$-$20.61 &$-$0.050 &$-$0.032\\
554 &$-$252.31 &160.29 &19.94 &E &$-$0.3$\pm1.6$ &0.9$\pm0.7$ &$-$21.09 &$-$0.049 &$-$0.018\\
560 &21.67 &170.53 &20.41 &E/S0 &$-$1.2$\pm2.1$ &1.5$\pm0.8$ &$-$20.62 &0.010 &0.029\\
572 &$-$198.72 &181.74 &20.29 &? &$-$3.1$\pm3.0$ &1.2$\pm0.9$ &$-$20.74 &$-$0.142 &$-$0.120\\
584 &$-$54.21 &201.30 &21.41 &? &$-$2.2$\pm9.9$ &0.6$\pm2.1$ &$-$19.62 &$-$0.453 &$-$0.461\\
626 &$-$56.50 &255.17 &20.32 &S0 &2.0$\pm4.5$ &1.2$\pm1.9$ &$-$20.71 &$-$0.089 &$-$0.068\\
1328 &$-$192.34 &$-$159.57 &21.77 &E &1.9$\pm5.7$ &1.4$\pm1.9$ &$-$19.26 &$-$0.044 &$-$0.062\\
1524 &82.16 &$-$39.65 &21.91 &S0 &$-$0.6$\pm4.4$ &1.0$\pm2.2$ &$-$19.12 &0.007 &$-$0.015\\
1842 &$-$14.42 &133.77 &22.27 &S0 &$-$4.2$\pm9.3$ &0.5$\pm3.5$ &$-$18.76 &0.130 &0.098\\
1865 &111.08 &146.35 &21.58 &? &$-$5.0$\pm10.7$ &0.3$\pm4.6$ &$-$19.45 &0.059 &0.046\\
\enddata
\tablenotetext{*}{Rest--frame absolute B magnitudes.}
\tablenotetext{\dag}{Radial color gradients in d(F606W--F814W)/dlogR, where R 
is the semi-major axis in arcsec.}
\tablenotetext{\ddag}{Radial color gradients corrected to a standard 
reference magnitude (using the best fit slope of Fig.~\ref{gradmag} of 
0.027).}
\end{deluxetable}

\clearpage
\begin{deluxetable}{crrrrrrrrrr}
\tablecaption{K+A Galaxies \label{tbl-2}}
\tablewidth{0pt}
\tablehead{
\colhead{Gal\#} & \colhead{x} & \colhead{y} & \colhead{V} & 
\colhead{Type} & \colhead{[OII]$\lambda$3727} & \colhead{$<$H$>$} & 
\colhead{$M_B$\tablenotemark{*}} & \colhead{CI\tablenotemark{\dag}} & 
\colhead{[CI]$_c$\tablenotemark{\ddag}}
\\
\colhead{} & \colhead{($\arcsec$)} & \colhead{($\arcsec$)} & \colhead{} & 
\colhead{} & \colhead{(\AA)} & \colhead{(\AA)} & \colhead{} & \colhead{}
}
\startdata
92 &137.23 &$-$237.13 &20.75 &S0 &0.0$\pm2.5$ &3.5$\pm1.7$ &$-$20.28 &$-$0.040 &$-$0.031\\
109 &73.68 &$-$222.07 &21.32 &S0 &$-$0.1$\pm4.3$ &7.0$\pm1.5$ &$-$19.71 &0.024 &0.018\\
167 &89.31 &$-$168.19 &21.55 &S0 &$-$3.6$\pm6.1$ &7.6$\pm2.3$ &$-$19.48 &0.010 &$-$0.002\\
200 &$-$12.63 &$-$138.44 &20.02 &S0 &$-$4.0$\pm5.9$ &3.4$\pm2.8$ &$-$21.01 &0.817 &0.846\\
206 &$-$165.28 &$-$135.64 &21.43 &? &$-$7.5$\pm7.6$ &3.8$\pm1.9$ &$-$19.60 &$-$0.114 &$-$0.123\\
209 &29.25 &$-$130.76 &20.09 &S0 &$-$0.8$\pm2.8$ &6.0$\pm0.8$ &$-$20.94 &0.015 &0.042\\
226 &$-$208.35 &$-$120.30 &21.31 &? &0.8$\pm3.5$ &3.7$\pm1.7$ &$-$19.72 &0.110 &0.104\\
243 &$-$13.78 &$-$99.71 &21.59 &S0 &2.1$\pm3.4$ &9.1$\pm0.9$ &$-$19.44 &0.075 &0.062\\
246 &78.11 &$-$96.07 &20.97 &S0 &3.4$\pm8.4$ &3.3$\pm1.9$ &$-$20.06 &0.004 &0.008\\   
290 &15.96 &$-$65.50 &21.16 &Irr &$-$6.9$\pm3.4$ &3.0$\pm1.9$ &$-$19.87 &0.127 &0.125\\
311 &$-$2.50 &$-$45.45 &21.32 &S0 &$-$5.8$\pm9.9$ &4.2$\pm3.5$ &$-$19.71 &$-$0.181 &$-$0.187\\
328 &$-$4.03 &$-$36.82 &20.11 &S0 &$-$1.0$\pm2.4$ &5.9$\pm1.1$ &$-$20.92 &$-$0.256 &$-$0.202\\
343 &$-$64.33 &$-$25.27 &20.80 &S0 &1.0$\pm5.0$ &6.2$\pm1.8$ &$-$20.23 &0.040 &0.048\\
344 &$-$38.25 &$-$24.87 &20.87 &S0 &$-$1.3$\pm4.6$ &3.3$\pm3.2$ &$-$20.16 &0.049 &0.055\\
346 &$-$8.17 &$-$23.07 &19.87 &S0 &$-$0.8$\pm2.4$ &3.0$\pm1.0$ &$-$21.16 &$-$0.131 &$-$0.098\\
354 &3.73 &$-$16.28 &21.11 &E &$-$4.9$\pm6.3$ &3.2$\pm2.2$ &$-$19.92 &$-$0.063 &$-$0.063\\
406 &$-$121.17 &24.15 &21.90 &? &$-$1.9$\pm6.3$ &3.1$\pm2.7$ &$-$19.13 &0.017 &$-$0.005\\
420 &$-$18.87 &36.69 &22.14 &S0 &1.0$\pm6.4$ &4.9$\pm2.0$ &$-$18.89 &0.153 &0.125\\
493 &11.71 &98.91 &21.70 &? &$-$1.2$\pm3.2$ &3.3$\pm1.7$ &$-$19.33 &$-$0.036 &$-$0.052\\
507 &$-$181.26 &112.42 &20.52 &S0 &$-$1.7$\pm2.5$ &4.5$\pm1.3$ &$-$20.51 &$-$0.338 &$-$0.322\\
562 &57.01 &175.13 &21.11 &E &0.6$\pm5.3$ &4.8$\pm2.7$ &$-$19.92 &0.341 &0.341\\
565 &140.21 &177.65 &21.86 &E &2.5$\pm5.2$ &4.9$\pm2.4$ &$-$19.17 &$-$0.098 &$-$0.119\\
594 &$-$169.55 &213.66 &21.90 &? &$-$9.4$\pm4.5$ &6.1$\pm1.3$ &$-$19.13 &0.021 &$-$0.001\\
1775 &116.86 &97.78 &21.74 &S0 &$-$0.1$\pm9.9$ &3.7$\pm4.5$ &$-$19.29 &$-$0.044 &$-$0.061\\
\enddata
\tablenotetext{*}{Rest--frame absolute B magnitudes.}
\tablenotetext{\dag}{Radial color gradients in d(F606W--F814W)/dlogR, where R 
is the semi-major axis in arcsec.}
\tablenotetext{\ddag}{Radial color gradients corrected to a standard 
reference magnitude (using the best fit slope of Fig.~\ref{gradmag} of 
0.027).}
\end{deluxetable}

\clearpage
\begin{deluxetable}{crrrrr}
\tablecaption{Results of the K-S Test\tablenotemark{a} \label{tbl-3}}
\tablewidth{0pt}
\tablehead{
\colhead{Sample} & \colhead{Raw} & \colhead{Min slope} & \colhead{Best Fit 
slope} & \colhead{Max slope}
}
\startdata
Entire Sample & 0.9\% & 1.0\% & 9.7\% & 14.6\% \\
Excluding \#200 & 1.8\% & 1.8\% & 17.3\% & 24.8\% \\
Excluding \#290 & 1.8\% & 1.8\% & 17.3\% & 24.8\% \\
Excluding \#507 & 0.6\% & 0.6\% & 7.7\% & 11.7\% \\
Excluding all 3 & 2.5\% & 2.1\% & 25.1\% & 34.6\% \\
\enddata
\tablenotetext{a}{Shown is the probability that the two samples are drawn 
from the same parent sample.}
\end{deluxetable}


\begin{references}

\reference{amb99} Abadi, M. G., Moore, B., \& Bower, R. G. 1999, \mnras, 308,
947

\reference{bal94} Balcells, M., \& Peletier, R. F. 1994, \aj, 107, 135

\reference{ba97} Balogh, M. L., Morris, S. L., Yee, H. K. C., Carlberg, 
R. G., \&  Ellingson, E. 1997, \apj, 488, L75

\reference{ba98} Balogh, M. L., Schade, D., Morris, S. L., Yee, H. K. C., 
Carlberg, R. G., \&  Ellingson, E. 1998, \apj, 504, L75

\reference{ba99} Balogh, M. L., Morris, S. L., Yee, H. K. C., Carlberg, 
R. G., \& Ellingson, E. 1999, \apj, 527, 54

\reference{b96} Barger, A. J., Arag\'{o}n-Salamanca, A., Ellis, R. S., Couch, 
W. J., Smail, I., \& Sharples, R. M. 1996, \mnras, 279, 1

\reference{bek99} Bekki, K. 1999, \apj, 510, L15

\reference{bra00} Bravo-Alfaro, H., Cayatte, V., van Gorkom, J. H., \&
Balkowski, C. 2000, \aj, 119, 580

\reference{bo78} Butcher, H., \& Oemler, A., Jr. 1978, \apj, 226, 559

\reference{bo84} Butcher, H., \& Oemler, A., Jr. 1984, \apj, 285, 426

\reference{crd} Caldwell, N., Rose, J. A., \& Dendy, K. 1999, \aj, 117, 140

\reference{cal96} Caldwell, N., Rose, J. A., Franx, M., \& Leonardi, A. 1996,
\aj, 111, 78

\reference{ca94} Cayatte, V., Kotanyi, C., Balkowski, C., \& van Gorkum, 
J. H. 1994, \aj, 107, 1003

\reference{crc00} Concannon, K. D., Rose, J. A., \& Caldwell, N. 2000, \apj, 
536, L19

\reference{cs87} Couch, W. J., \& Sharples, R. M. 1987, \mnras, 229, 423

\reference{cou98} Couch, W. J., Barger, A. J., Smail, I., Ellis, R. S., \&
Sharples, R. M. 1998, \apj, 497, 188

\reference{dg83} Dressler, A., \& Gunn, J. E. 1983, \apj, 270, 7

\reference{ad94} Dressler, A., Oemler, A., Jr., Sparks, W. B., \& Lucas, R. A.
1994, \apj, 435, L23

\reference{ad97} Dressler, A., Oemler, A., Jr., Couch, W. J., Smail, I., 
Ellis, R. S., Barger, A., Butcher, H., Poggianti, B. M., \& Sharples, R. M. 
1997. \apj, 490, 577

\reference{ad99} Dressler, A., Smail, I., Poggianti, B. M., Butcher, H.,
Couch, W. J., Ellis, R. S., \& Oemler, A., Jr. 1999, \apjs, 122, 51

\reference{fmb91} Fabricant, D. G., McClintock, J. E., \& Bautz, M. W. 1991,
\apj, 381, 33

\reference{fa00} Fasano, G., Poggianti, B. M., Couch, W. J., Bettoni, D.,
Kj$\ae$rgaard, P., \& Moles, M. 2000, \apj, 542, 673

\reference{f98} Fisher, D., Fabricant, D., Franx, M., \& van Dokkum, P. 1998,
\apj, 498, 195

\reference{f93} Franx, M. 1993, \apj, 407, L5

\reference{ga95} Gavazzi, G., Contursi, A., Carrasco, L., Boselli, A., 
Kennicutt, R., Scodeggio, M., \& Jaffe, W. 1995, A\&A, 304, 325

\reference{gg72} Gunn, J. E., \& Gott, J. R. 1972, \apj, 176, 1

\reference{jan00} Jansen, R. A., Franx, M., Fabricant, D., \& Caldwell, N. 
2000, \apjs, 126, 271

\reference{kk99} Kenney, J. D. P., \& Koopmann, R. A. 1999, \aj, 117, 181 

\reference{lh88} Lavery, R. J., \& Henry, J. P. 1988, \apj, 330, 596

\reference{lpm92} Lavery, R. J., Pierce, M. J., \& McClure, R. D. 1992, \aj,
104, 2067

\reference{mh96} Mihos, J. C., \& Hernquist, L. 1996, \apj, 464, 641

\reference{moo96} Moore, B., Katz, N., Lake, G., Dressler, A., \& Oemler, A.,
Jr. 1996, Nature, 379, 613

\reference{mlk98} Moore, B., Lake, G., \& Katz, N. 1998, \apj, 495, 139

\reference{mw93} Moss, C., \& Whittle, M. 1993, \apj, 407, L17

\reference{mw00} Moss, C., \& Whittle, M. 2000, \mnras, 317, 667

\reference{new90} Newberry, M. V., Boroson, T. A., \& Kirshner, R. P. 1990,
\apj, 350, 585

\reference{nor01} Norton, S. A., Gebhardt, K., Zabludoff, A. I., \& Zaritsky, 
D. \apj, 557, 150

\reference{pel90} Peletier, R. F., Davies, R. L., Illingworth, G. D., Davis,
L. E., \& Cawson, M. 1990, \aj, 100, 1091

\reference{po99} Poggianti, B. M., Smail, I., Dressler, A., Couch, W. J.,
Barger, A., Butcher, H., Ellis, R. S., \& Oemler, A., Jr. 1999, \apj, 518, 
576

\reference{ro01} Rose, J. A., Gaba, A. E., Caldwell, N., \& Chaboyer, B. 
2001, \aj, 121, 793

\reference{sil89} Silva, D. R., Boroson, T. A., Thompson, I. B., \& 
Jedrzejewski, R. I. 1989, \aj, 98, 131

\reference{sol01} Solanes, J. M., Manrique, A., Garcia-Gomez, C., 
Gonzalez-Casado, G., Giovanelli, R., \& Haynes, M. P. 2001, \apj, 548, 97

\reference{th88} Thompson, L. 1988, \apj, 324, 112

\reference{vad88} Vader, J. P., Vigroux, L., Lachi\`{e}ze-Rey, M., \& 
Souviron, J. 1988, A\&A, 203, 217

\reference{vd98} van Dokkum, P. G., Franx, M., Kelson, D. D., Illingworth, G. 
D., Fisher, D., \& Fabricant, D. 1998, \apj, 500, 714

\reference{vd99} van Dokkum, P. G., Franx, M., Fabricant, D., Kelson, D. D., 
Illingworth, G. D. 1999, \apj, 520, L95

\reference{zab96} Zabludoff, A. I., Zaritsky, D., Lin, H., Tucker, D., 
Hashimoto, Y., Shectman, S. A., Oemler, A., \& Kirshner, R. P. 1996, \apj, 
466, 104

\end{references}
\end{document}